\documentclass[%
  prx,longbibliography,twocolumn,
showpacs,
 amsmath,amssymb,
 aps,
showkeys
]{revtex4-1}
\usepackage[utf8]{inputenc}
\usepackage[english]{babel}
\usepackage{graphicx}%
\usepackage[dvipsnames]{xcolor}
\usepackage{dcolumn}%
\usepackage{bm}%
\usepackage{footnote}
\usepackage{color}
\usepackage{hyperref}%

\hypersetup{
    colorlinks=true,
    linkcolor=blue,
    urlcolor=blue,
    citecolor=blue,
} %

\usepackage[normalem]{ulem} %

 \usepackage{epsfig}
 \usepackage{color}
 \usepackage{amsmath}
 \usepackage{amsthm}
 \usepackage{amsfonts}
 \usepackage{amssymb}
 \usepackage{tikz}
 \usepackage{dsfont} %

\def\be{\begin{equation}}
\def\eea{\end{eqnarray}}
\def\ee{\end{equation}}
\def\bea{\begin{eqnarray}}
\def\ea{\end{array}}
\def\ba{\begin{array}}

\def\zzz{{\mathchoice {\hbox{$\sf\textstyle Z\kern-0.4em Z$}}
{\hbox{$\sf\scriptstyle Z\kern-0.3em Z$}}
{\hbox{$\sf\scriptscriptstyle Z\kern-0.2em Z$}}
{\hbox{$\sf\textstyle Z\kern-0.4em Z$}}}}

\emergencystretch=\maxdimen
\usepackage{lipsum}
\begin{document}

\title{
What does it take to solve the 3D Ising model?\\
Minimal necessary conditions for a valid solution}

\author{
  G. M. Viswanathan$^{1}$, M. A. G. Portillo$^{1,2}$,
  E. P. Raposo$^3$,  M. G. E. da Luz$^2$}

\affiliation{
  $^1$Department of  Physics and
National Institute of Science and Technology of Complex Systems,
Federal University of  Rio Grande do Norte, Natal--RN, 59078-970,
Brazil\\
  $^2$Departmento de F\'{\i}sica, Universidade Federal do Paran\'a,
Curitiba--PR, 81531-980, Brazil \\
  $^3$Laboratório de Física Teórica e Computacional, Departamento de
Física, Universidade Federal de Pernambuco, Recife--PE, 50670-901,
Brazil  
}
\date{\today}%

\begin{abstract}
Exact solution of the Ising model on the simple cubic lattice is
one of the long-standing open problems in rigorous statistical mechanics.
{Indeed, it is generally believed that 
  settling it would constitute a methodological breakthrough,
  fomenting great prospects for further application, similarly to
  what happened when Lars Onsager solved the  two dimensional  model 
  eighty years ago.}
Hence, there have been many attempts to find analytic expressions
for the exact partition function $Z$, but all such attempts
have
failed due to unavoidable conceptual or mathematical obstructions.
Given the importance of this simple yet
  paradigmatic model,
here we set out clear-cut criteria for any claimed exact
expression for $Z$ to be minimally plausible.
Specifically, we present six necessary --- but not sufficient ---
conditions that $Z$ must satisfy.
These criteria {will} allow very quick
plausibility checks of future claims.
As illustrative examples, we discuss previous {mistaken
``solutions,''} unveiling their shortcomings.

\end{abstract}

\maketitle

\section{Introduction}

Exact results are always welcome in science, even if
simplified or idealized models of more realistic natural
phenomena \cite{Gershenfeld-1998}.
For example, an elucidating discussion about the general
importance of analytical solutions in physics can be
found in Ref. \cite{borwein-2013}.
The Ising model of magnetism was originally proposed by Wilhelm Lenz
in 1920 and exactly solved~\cite{ising1924} in one dimension by his
graduate student Ernst Ising in 1924.  The Ising model long ago
ceased to be a paradigm restricted only to magnetism models. 
Currently it has found applications in diverse areas, from
neuroscience to sea ice and voter models, to name a
few~\cite{tranquillo}.
Since Lars Onsager's solution of the 2D Ising model 
in 1942 (which was published in 1944~\cite{ons})
statistical physics in general and equilibrium statistical
mechanics in particular have experienced a great flourishing
of powerful mathematical techniques
\cite{ruelle-1999,thompson-2015}, allowing considerable
progress towards obtaining exact expressions for many
relevant models \cite{sutherland-2005,er1,er2}.

In spite of all these advances, one of the most paradigmatic
systems in statistical physics, namely the ferromagnetic Ising model
 with interactions between nearest-neighbor two-state spins 
 on the simple cubic
lattice~\cite{singh-2020} --- henceforth referred as the 3D
Ising model --- has withstood all attempts at exact solution.
According to Rowlinson~\cite{maddox2}, the first claim of
analytically cracking the 3D Ising model was presented at StatPhys
2, held in Paris in 1952, by John R. Maddox, who later became editor
of {\it Nature} (see also Refs.
~\cite{maddox1,maddox3,perk-comment}). Eliott W. Montroll, still
during the conference, showed that the proposed expression could not
be correct, by comparing it with the first few terms of the well
established exact series expansions for the high and low temperature
limit cases. It was identified later that the error was due to an
incorrect application of the Jordan-Wigner transformation. Since
then, new announcements of an exact solution have been made every
few years, only to be systematically proved incorrect (e.g.,
Refs. \cite{das,lou,zd-zhang,degang}).  What is seemingly lacking,
in this historical context, is a set of clear-cut plausibility
criteria that can be used to quickly verify whether or not a claimed
solution is minimally worth considering. Therefore, we shall list
a number of necessary --- but not sufficient --- conditions that a
correct partition function $Z$ must satisfy. Although some of
these have already (in part or in full) appeared in the literature,
here they are presented in an unified, rigorous and comprehensive
way.  {We emphasize} that on the one hand these conditions can
be used to refute unfounded claims --- and thereby to identify any
inappropriate protocols that are responsible for the incorrect
features for $Z$. On the other hand, they might serve as a guide for
the development of promising solution schemes.  {Finally, we
  briefly mention two reasons why the 3D Ising model is considered a
  {\em major} open problem, besides the obvious intellectually
  instigating fact that it still stands eighty years since Onsager
  solved the 2D version. First, despite the simple definitions 
  of Ising-like models, their usefulness in studying a large number
  of diverse processes is overwhelming
  \cite{singh-2020,majewski-2001,binek-2003,suzuki-2013,
    mccoy-2014,adler-2016,fadil-2020,farah-2020,lipowski-2022}. But
  this should not be a surprise. Indeed, given the universality
  classes of classical spin models, of which the Ising is probably
  the most emblematic example, they all can be mapped to distinct
  instances of the logic problem of satisfiability (SAT)
  \cite{cuevas-2016}. In this way, the analytic solution of the 3D
  case conceivably would represent a great boost for its already
  wide applicability. Second, the 3D Ising model near the critical
  point is very closely related to string and gauge field 
  theories \cite{polyakov-1987}. In particular, these associations
  can be analyzed through conformal bootstrap methods
  \cite{showk-2012} (for a review see, e.g., \cite{rychkov-2020}).
  Hence, the eventual determination (or eventual disproof of
  existence) of an exact solution certainly will impact other
  important areas of physics.}

{All these potential perspectives
  involving a rigorous analytical expression for the partition
  function of the 3D Ising model certainly justify the
  establishment of a minimal set of clear-cut criteria for the
  validation of its exact solution, as presented below.}

\section{The 3D Ising model}

Suppose a finite simple cubic lattice ${\mathbb Z}^3_N$
of size $L \times L \times L$, with $L$ a finite
positive integer. 
There are $N = L^3$ sites that can be tagged as ${\bf l} = 
(l_1, l_2, l_3)$ for $i=1,2,3$ indicating
the $i$-th  spatial direction  (with  unit vector 
$\widehat{e}_i$) and $l_i = 1, \ldots, L$.
Consider also two disjoint lattices, 
$\widetilde{{\mathbb Z}^3_N}$
and $\partial {\mathbb Z}^3_N$, of sites labeled, 
respectively, by ${\bf l}_{b} = (n_1, n_2, n_3)$,
with $2 \leq n_i \leq L-1$,
and  ${\bf l}_{f} = (m_1, m_2, m_3)$, 
where at least one $m_i$ is necessarily 1 or $L$.
Note that ${\mathbb Z}^3_N = 
\widetilde{{\mathbb Z}^3_N} \cup
\partial {\mathbb Z}^3_N$, for 
$\widetilde{{\mathbb Z}^3_N}$
($\partial {\mathbb Z}^3_N$)
representing the ``bulk" (``frontier" or boundary) sites of
${\mathbb Z}^3_N$.
In other words, $\partial {\mathbb Z}^3_N$ is the boundary
or surface whereas $\widetilde{{\mathbb Z}^3_N}$ is the bulk
or interior of the lattice. 

The spin variable $\sigma_{\bf l}$ at the vertex 
${\bf l} \in {\mathbb Z}^3_N$ can assume only two possible 
values, namely,  -1 and +1.
The Hamiltonian of the 3D anisotropic 
Ising model on ${\mathbb Z}^3_N$ and with zero external 
magnetic field is given by 
\begin{eqnarray}
H_{N}(\sigma) &=& - 
\sum_{i=1}^{3}  J_i \,  \bigg(
\sum_{{\bf l}_b'' - {\bf l}_b' = \hat{e}_i}
\sigma_{{\bf l}_b''} \,
\sigma_{{\bf l}_b'} 
+ \!
\sum_{{\bf l} - {\bf l}_f = 
\varepsilon({\bf l}) \, \hat{e}_i}
\sigma_{{\bf l}} \,
\sigma_{{\bf l}_f} \bigg),
\nonumber \\
&=& H_{N}^{(b)}(\sigma) + 
H_{N}^{(f)}(\sigma)~,
\label{eq:energy}
\end{eqnarray}
where $\varepsilon({\bf l}) = \pm 1$ if 
${\bf l} \in  \widetilde{{\mathbb Z}^3_N}$ and 
$\varepsilon({\bf l}) = + 1$ if 
${\bf l} \in  \partial {\mathbb Z}^3_N$.
The quantities $J_{1,2,3}$ are the couplings constants in the
three distinct spatial directions $i=1,2,3$.
The two terms in the last equality in Eq.  (\ref{eq:energy})
can readily be identified as the energy contributions from
the system bulk $b$ and frontier $f$ regions for a state
$\sigma$ of the system.
In fact, each $\sigma$
represents a possible distinct configuration of -1's and +1's
along the sites of the whole lattice and characterizes a
specify system state.

Different boundary conditions (BCs) can be imposed to
the problem \cite{b-c-1,b-c-2}
They essentially specify constraints on the spin
configuration of the set $\{\sigma \}_{\partial {\mathbb Z}^3_N}$
(see below).
       
The canonical partition function is conventionally
defined as $Z_N(K_1,K_2,K_3) = \sum_{\sigma}
\exp[\beta \, H_N(\sigma)]$,
where $\beta = (k_B \, T)^{-1}$ and  $K_i = \beta \,
J_i$ is the $i$-th reduced temperature.
The sum is over all the possible spin
configurations $\{\sigma\}$ over ${\mathbb Z}^3_N$
(observing the specified BCs).
The partition function per site in the thermodynamic
limit, the object of our interest here, is defined as
\be
Z(K_1,K_2,K_3)
=  \lim_{N\to \infty} 
\big[Z_N(K_1,K_2,K_3)\big ]^{1/N}~.
\label{eq-Ising3d}
\ee
The challenge of finding an exact analytic expression for
the above $Z$ has been called the ``holy grail of statistical
mechanics''~\cite{cipra-2000}.
To this day it remains one of the most important
unsolved problems in statistical physics.

\section{Necessary conditions for a valid solution}

Our main goal in this contribution is to establish a set
of six necessary conditions that must be satisfied by any
prospective of an 3D Ising model exact analytic expression
for $Z(K_1, K_2, K_3)$.
Although the list consists of necessary conditions, these are not also
sufficient conditions.  In other words, any claimed $Z$ may be wrong
even if all necessary conditions are satisfied.  However, if even just
one condition is violated, the claimed expression for $Z$ is 
certainly wrong.

Next we present the mentioned conditions in a order
somehow going from the more basic and fundamental to the more
technical and abstract.

\subsubsection{Condition 1}

{\it In the thermodynamic limit the per-site partition function
  $Z$ of the 3D Ising model must be independent of the boundary
  conditions}
\cite{ruelle-1974, velenik-2017,perk-comment}.

Surprisingly, the very broad reaching of this requirement seems not to
be properly appreciated as one should expect.  Some recent erroneous
claims of an exact $Z$ even have assumed that certain special BCs
could violate the condition 1 (see Sec. \ref{sec:examples}).  Given
such misunderstandings, below we present a very general and rigorous
(although concise) proof that this indeed must be the case.

The thermodynamic limit represented by Eq. (\ref{eq-Ising3d})
is known to be equivalently stated in terms of sequences of
subgraphs $G_k$ of $\mathbb{Z}^3$ (e.g., see
\cite{feynman-page149,gattringer}).
There is a large relative freedom in choosing the structure of
the successive $G_k$'s, provided they satisfy three fundamental
properties, known as Van Hove's assumptions 
\cite{velenik-2017}.

Denoting the number of vertices of a finite lattice $G$ as $V(G)$,
these assumptions are as follows:
\begin{itemize}
    \item $\cup_k \, G_k = \mathbb{Z}^3$,
    \item $G_{k} \subset G_{k+1}$,
    \item  $\lim_{k \rightarrow \infty}
    { V(\partial G_k)}/{V(G_k)}=0$, 
    for $\partial G_k$ representing the frontier
    of $G_k$, namely, \\
    $\partial G_k =
    \{ {\bf l} \in G_k \, | \, \exists  \, {\bf j} 
    \notin G_k, \, |{\bf l} - {\bf j}|=1\}$.
\end{itemize}

From the above it is also possible to define
$\widetilde{G_k} = \{{\bf l} \in G_k \, | \, {\bf l} 
\notin \partial G_k\}$, which is the bulk or core graph 
associated to $G_k$.
As a trivial example of a $G_k$ satisfying all the above
characteristics, we mention the previously mentioned
limited cubic lattice ${\mathbb Z}^3_N$,
where $k = N = L^3$.

Thus, $Z(K_1, K_2, K_3)$ can be written more generally as
\begin{equation}
  Z(K_1,K_2,K_3) = c \,
  \lim_{k \rightarrow \infty}\! 
  \bigg[Z_{G_k}(K_1,K_2,K_3)\bigg]^{1\over V(G_k)},
\label{eq:subgraph}
\end{equation}
where 
$$Z_{G_k}(K_1,K_2,K_3) =
\sum_{\sigma \in \{\sigma\}_{G_k}} \exp[\beta \, H_{G_k}(\sigma)]$$
and $H_{G_k}(\sigma)$ is the natural extension of Eq. (\ref{eq:energy})
to $G_k$.
{
Further, $c$ is a constant of a purely topological origin.
It may differ from 1 depending on the characteristics of
the chosen sequence $G_k$.
However, it should not alter the resulting physics associated
to the obtained partition function.
Therefore, $c$ might be set equal to 1  for sake of
discussion.}

An important aspect of the finite Ising model relates
to the boundary conditions assumed for the $G_k$'s.
A rather general formulation for typical 
BCs relies on the following construction.
Let $\Omega(G)$ represent {\em all the possible combinations}
of the spin configurations on the vertices of the finite $G$,
i.e., an element of $\Omega$ is denoted by $\sigma$ and 
is a map $\sigma: G \rightarrow \{-1,1\}$.
Consider then $\partial G_k$  and a specific subset 
$\Gamma_{BC}^k \subset \Omega(\partial G_k)$.
We say that $\Gamma_{BC}^k$ determines the BCs on
the Ising model if the allowed spin configurations
$\sigma_{BC}$ belong to
$$\Omega_{BC}(G_k) =
\Omega(\widetilde{G_k}) \times \Gamma_{BC}^k.$$
For instance, for $\Gamma_{BC}^k = \Omega(\partial G_k)$ 
we have the usual free BCs, namely, for any site
in the frontier the spin value can assume both values,
-1 and +1, without restrictions.
On the other hand, for $G_k$ displaying torus (or
periodic), cylindrical, Klein, twisted, screw, etc, 
topology, then the permitted configurations in 
$\Gamma_{BC}^k$ are established by proper pairwise
mappings in the form 
$\sigma_{{\bf l}_f''} \leftrightarrow 
\sigma_{{\bf l}_f'}$.
Hence, the finite partition function with 
the BCs determined by $\Gamma_{BC}^k$ reads 
\begin{equation}
{Z_{G_k}^{BC}}(K_1,K_2,K_3) =
\sum_{\sigma \in {\Omega_{BC}(G_k)}} \! \! \!
\exp[\beta (H_{G_k}^{(b)}(\sigma)
  + H_{G_k}^{(f)}(\sigma))].
\label{eq:hfactor}
\end{equation}

We can now easily show that in the proper limit the
partition function is independent on the BCs.
We first observe that for the free BCs
\begin{equation}
\lim_{k \rightarrow \infty}
\frac{\ln[Z^{\mbox{\scriptsize free}}_{G_k}]}{V(G_k)}
\label{eq:limit-free}
\end{equation}
exists (for a proof see Ref. \cite{velenik-2017}).
Let us denote the limit in Eq. (\ref{eq:limit-free}) 
as $\ln[Z(K_1,K_2,K_3)]$.
Second, we determine a bound for $H_{G_k}^{(f)}(\sigma)$ 
in Eq. (\ref{eq:hfactor}).
For any site ${\bf l}_f$ in $\partial G_k$, 
the maximum number of ${\bf l} \in G_k$ such that 
$|{\bf l} - {\bf l}_f| = 1$ is 5 for the simple cubic lattice, then we have
that $|H_{G_k}^{(f)}| < 5 \, V(\partial G_k)$.
Third, by lemma 2.2.1 in \cite{ruelle-1974}
\begin{equation}
\bigg| \ln[Z^{BC}_{G_k}] - 
\ln[Z^{\mbox{\scriptsize free}}_{G_k}] \bigg| 
\leq \beta \, |H^{(f) \, BC}_{G_k}-H^{(f) \,
\mbox{\scriptsize free}}_{G_k}|.
\label{eq:free-diferenca}
\end{equation}
Lastly, by dividing Eq. (\ref{eq:free-diferenca}) 
by $V(G_k)$, considering the triangular inequality
$$|H^{(f) \, BC}_{G_k}-H^{(f) \,
\mbox{\scriptsize free}}_{G_k}|
\leq 2 \times \beta \, 5 \, V(\partial G_k)$$
and since the sequence $G_k$ is van Hove, then
\begin{equation}
 \lim_{k \rightarrow \infty}
   \frac{\ln[Z^{BC}_{G_k}]}{V(G_k)} =
    \lim_{k \rightarrow \infty}
    \frac{\ln[Z^{\mbox{\scriptsize free}}_{G_k}]}{V(G_k)}.
    \label{eq:reducedlimit}
\end{equation}
Thus, we readily conclude that
\begin{equation}
  \lim_{k \rightarrow \infty}
\left(
Z_{G_k}^{BC}(K_1,K_2,K_3)\right)^{1/V(G_k)}
=
Z(K_1,K_2,K_3)
\label{eq:zbc}
\end{equation}
is well defined and independent on the BCs.

\subsubsection{Condition 2}
{\it
The claimed partition function per site of the 3D model
$Z(K_1,K_2,K_3)$ must reduce to Onsager's solution 
whenever one of the three reduced temperatures vanishes.
}

Indeed, suppose without loss of generality that $J_3=0$,
so that $K_3=0$. 
Then, for ${\bf l}_l$ indicating the sites which lie in
the plane $x_3 = l$ (whose set of spin configurations we 
represent  by $\{\sigma^l\}$), the 3D Hamiltonian of size 
$N = L\times L\times L = L^2 \times L$ 
can be written as ($\sigma_{{\bf j}} \equiv 0$ if
${\bf j} \notin G_k$)
\begin{equation}
H_{L^2 \times L}(\sigma) = - \sum_{l=1}^{L} \, 
\sum_{{\bf l}_l} \, \sum_{i=1}^{2}  
J_i \, \sigma_{{\bf l}_l} \, 
\sigma_{{\bf l}_l +{\hat e}_i} =
\sum_{l=1}^{L} H_{L^2}(\sigma^l),
\end{equation}
with $H_{L^2}(\sigma^l)$ the energy associated to the plane
$x_3=l$ for $\sigma^l$ a given distribution of spins in such
plane.
Note that the 3D Hamiltonian is now expressed as the sum of
$L$ independent and identical 2D Hamiltonians.
In this case, the 3D partition function
$Z_N = Z_{L^3}(K_1,K_2,0) = Z_{L^2 \times L}$
factors as
\begin{equation}
Z_{L^2 \times L} = \sum_{\sigma^1} \ldots
\sum_{\sigma^L } \, \prod_{l=1}^{L}
\exp[-\beta \, H_{L^2}(\sigma^l)] = (Z_{L^2})^L,
\label{eq-eoirjwoirjtnonjojo4i5joijoi4ntb}
\end{equation}
where
$Z_{L^2} = Z_{L^2}(K_1,K_2)$ is the finite 2D partition function,
whose per-site thermodynamic limit $Z(K_1,K_2)$ is naturally
defined as
\be
Z(K_1,K_2)
=  \lim_{L\to \infty} 
\big[Z_{L^2}(K_1,K_2)\big ]^{1/L^2}~.
\ee
Thus, from Eq. (\ref{eq-eoirjwoirjtnonjojo4i5joijoi4ntb})
we get
\bea
Z(K_1,K_2,0) &=& \lim_{L\to \infty}
Z^{1/L^3}_{L^2 \times L} = \lim_{L\to \infty} (Z^L_{L^2})^{1/L^3}
\nonumber \\
&=&
\lim_{L\to \infty} Z^{1/L^2}_{L^2} = Z(K_1,K_2).
\eea
Thus, the 3D per-site partition function must reduce to
Onsager's solution when any of the 3 couplings is
made to vanish.

\subsubsection{Condition 3}

{\it For the isotropic case, namely $K = K_i$
    ($i=1,2,3$), any claimed per-site partition function must be
    analytic for $0 \leq  K < K_c$ \cite{ott_2020} and for
    {$K_c < K < \infty$ \cite{ott_2020_preprint}},
    where $K_c = 0.221\,654\,626...$ is the well known
numerically estimated value of the 
    critical
    temperature of the 3D Ising model.}

Of course, when there are interactions besides nearest-neighbor
or when there is an applied magnetic field, the range of
analyticity in $K$ for the partition function can be distinct
from the above one (see, for instance, refs.
\cite{ott_2020,ott_2020_preprint,velenik-2017}).

Note also that an exact expression for $Z$ should lead to an
exact formula for $K_c$.
Thence, if one has derived a tentative exact $Z$, its
analyticity should be relatively easy to test, e.g.,
for complex functions from the Cauchy-Riemann equations
\cite{mardsen-1998}, and for real functions using standard
techniques, as those described in \cite{krantz-2002}.
We observe that the numerical value of the critical 
temperature is known with very high precision from
Monte Carlo simulations and other numerical approaches 
\cite{beta-c,blote-1996,landau2018,perk-comment2}.

\subsubsection{Condition 4} 

\begin{table}[t!] 
\caption{The first ten nonzero coefficients in the high 
temperature expansion of
  ${\mathcal Z}_{\mbox{\tiny high}}(v) = 
  \sum_{n=0}^\infty a_n v^n$ and in the low 
temperature expansion of
${\mathcal Z}_{\mbox{\tiny low}}(u)= \sum_{n=0}^\infty 
b_n u^n$ (see main text). 
  The  coefficients 
  were obtained by Guttmann and Enting using the finite lattice
  method~\cite{guttmann-enting}.}
\begin{tabular}{crc||cr}
\hline
  $n$ (high) & $a_n$ & &  \ \ \ \ $n$ (low)  \ \ \ \ & $b_n$ \\
\hline
0  & 1          & & 0  & 1      \\
4  & 3          & & 3  & 1      \\
6  & 22         & & 5  & 3      \\
8  & 192        & & 6  & -3     \\
10 & 2046       & & 7  & 15     \\
12 & 24853      & & 8  & -30    \\
14 & 329334     & & 9  & 101    \\
16 & 4649601    & & 10 & -261   \\
18 & 68884356   & & 11 & 807    \\
20 & 1059830112 & & 12 & -2308  \\
\hline
\end{tabular}
\label{table1}
\end{table}

{\it For the isotropic case, the series expansion in the high (low)
  temperature limit --- $K$ small (large) --- of the claimed solution
  must perfectly match the already established series to all known
  orders (see, e.g., Refs. \cite{ott_2020,velenik-2017}).}

This is a direct consequence of the fact that in
the domain of an analytic function, its series expansion around
a fixed expansion point must be unique.

Let $v_i = \tanh[K_i]$ be the high temperature 
expansion variable.
For the isotropic case, i.e., $v = v_i$ ($i=1,2,3)$, 
we define
\be
Z(K)= 2 \, \cosh[K]^3 \, 
\mathcal Z_{\mbox{\tiny high}}(v).
\ee
The first few expansion terms of
$\mathcal Z_{\mbox{\tiny high}}(v)$ have been rigorously 
determined \cite{velenik-2017} via finite lattice 
methods. 
From the Condition 3, $Z(K)$ is analytic for 
$T > T_c$.
Therefore, for large $T$ the function 
$\mathcal Z_{\mbox{\tiny high}}(v)$ is also analytic.
In this way, the series expansion of 
$Z(K) / (2 \, \cosh[K]^3)$ must coincide 
with the mentioned known terms.

Likewise, by setting $u_i = \exp[-4 K_i]$ as the low
temperature variable, for the isotropic case we have 
$u_i=u$ ($i=1,2,3$). 
By writing the partition function as
\be
Z(K)=  u^{-3/4} \, \mathcal Z_{\mbox{\tiny low}}(u),
\ee
the first few exact terms of the low temperature expansion 
of $\mathcal Z_{\mbox{\tiny low}}(u)$ has also being
calculated \cite{guttmann-enting}.
So, similarly to the high temperature expansion,
for $T <T_c$  low enough, any claimed exact solution 
should meet term-by-term the mentioned series.

The first few expansion terms of both $\mathcal Z_{\mbox{\tiny
    high}}(v)$ and $Z_{\mbox{\tiny low}}(u)$ are known, thanks to
finite lattice method \cite{guttmann-enting}.  Ian Enting and Tom de
Neef originally developed this innovative approach in the 1970s for
generating series expansions, with applications to exact enumeration
problems \cite{enting-de-neef}.  Since then, this powerful numerical
method has led to significant advances.  For example, Iwan Jensen has
used it to calculate the statistics of self-avoiding polygons on the
square lattice~\cite{jensen2000}.  Very recently, Nathan Clisby has
written an expository article concerning the method's relevance to the
enumerative combinatorics of lattice polymers~\cite{clisby}.  It is
arguably the most powerful algorithmic technique for obtaining exact
series expansions to high order of models whose exact solution is not
known, including of course the 3D Ising model.

Table~\ref{table1} lists the first few non-zero terms
of the high and low temperature expansions.
So, according to the present Condition, any claimed $Z$
which does not exactly comply with the terms in 
Table~\ref{table1} cannot be the exact partition function
for the 3D Ising model.

\subsubsection{Condition 5}

{\it 
The claimed exact $Z$ should display permutation symmetry and 
convexity on the reduced temperature variables
$K_i = \beta \, J_i$.}

The first property above, as proposed in \cite{perk-comment},
implies that $Z(K_1,K_2,K_3)=Z(K_{\pi(1)},K_{\pi(2)},K_{\pi(3)})$
where $\pi(i)$ is a permutation for $i=1,2,3$. 
Since the final $Z$ does not depend on 
the BCs or on the specific choice of the van Hove sequence, 
consider $G_k$ in Eq. (\ref{eq:subgraph}) as the 
simple cubic lattice ${\mathbb Z}^3_{N}$ with free
BCs. 
Then, the proof is direct since the finite
$Z_N(K_1,K_2,K_3)$ trivially presents the aforementioned
symmetry.
Further (refer to Ref. \cite{perk-comment2}),
if one can use $Z$ to derive an equation for the 
critical $\beta_c$, i.e., 
$F(\beta_c \, J_1, \beta_c \, J_2, \beta_c \, J_3)=0$, 
then also $\beta_c$ must be invariant under any 
permutation of $J_1$, $J_2$ and $J_3$.

In  Ref. \cite{velenik-2017} it has been shown that
for the isotropic case $J_i = J$ ($i=1,2,3)$, the 
partition function must be a convex function in $\beta$.
For the anisotropic case, the proof follows exactly the
same steps.
Indeed, considering again $G_k$ the finite cubic lattice
with free BCs, one finds that $G_k$ is convex (details
omitted here).
But since the limit of a van Hove sequence of convex
functions is also convex \cite{niculescu-2006}, the final
result holds.
As a consequence, one has the following.
For $\alpha \in [0,1]$, $\beta_a, \beta_b \in (0,\infty)$ 
and arbitrary $J_1, J_2, J_3$, then for
$\beta_d = \alpha \, \beta_a + (1-\alpha) \, \beta_b$
\begin{eqnarray}
Z(\beta_d \, J_1, \beta_d  \, J_2, \beta_d \, J_3)
&\leq& \alpha \, Z(\beta_a \, J_1, \beta_a \, J_2, 
\beta_a \, J_3)
\nonumber \\
& & + (1 - \alpha) \, Z(\beta_b \, J_1, \beta_b \, J_2,
\beta_b \, J_3).
\nonumber \\
\end{eqnarray}

\subsubsection{Condition 6}
{
\it
The claimed exact solution must bring clarity to the conundrum
related to the behavior of the partially resummed high
temperature expansion of the anisotropic partition function
\cite{hansel,guttmann-prl}.
}

Based on the  anomalous behavior of the partially resummed
series solution of the 3D Ising model, it is now believed
that the 3D Ising model might ``not be solvable''
\cite{guttmann-prl}, by which is
meant that $Z$ may not be differentiably
finite (i.e., $D$-finite). In other words, the solution
is not an holonomic function.
Recall that any function that is analytic and satisfies a
linear differential equation with polynomial coefficients
is said to be holonomic (see, e.g., \cite{hoven-2001}).

Hence, any claimed  exact holonomic $Z$ must be able 
to explain how and why the anisotropic 3D Ising model
has a high-temperature series which, upon partial 
resummation, seems to indicate non-$D$-finiteness.  

\section{Some Previously Claimed Solutions in the Literature
\label{sec:examples}}

We briefly review how previous advanced solutions have 
failed to satisfy the above conditions, thus not 
representing the correct exact $Z(K_1, K_2, K_3)$.

As already mentioned, the proposal by Maddox in 1952 
violated the Condition 4 for the series expansions, 
as did the ones by Das~\cite{das}, Lou and 
Wu \cite{lou}, and Z.-D.~Zhang ~\cite{zd-zhang}. 
The serious errors in the latter has also been extensively 
addressed in Refs. 
\cite{fisher-comment,perk-comment}.
Moreover, except for the solution proposed by Zhang, all others
also violate Condition 3.
In fact, Zhang's solution only seems to satisfy it because
the numerical value of the critical temperature is imposed as
an ansatz, built into his construction.
Still, Zhang's critical temperature of $K_c \approx$
\mbox{0.221~658~637~208~698}~\cite{zd-zhang} (taken from a conjecture of
Rosengren~\cite{rosengren}) differs from the best known
numerical estimate $K_c\approx$ 0.221~654~626.

Crucially, none of the above claimed solutions minimally attend
Condition 6.
As explained above, the resummed high temperature series of the
anisotropic 3D model seems to show an anomalous behavior,
strongly suggesting non-$D$-finiteness.
But the claimed solutions all behave normally under resummation
of the anisotropic high temperature series --- a glaring
discrepancy.

In 2021, D.~Zhang~\cite{degang} (not to be confused with Z.-D.~Zhang)
made another claim, promptly criticized in \cite{perk-comment2}.  It
is easy to check that the assertion in \cite{degang} violates
Conditions 1 and 4. The claimed solution also fails to bring new
insights regarding condition 6.  For condition 3, the predicted
critical temperature disagrees drastically with the known numerically
estimated value.  Finally, regarding Condition 2, Zhang writes
(inaccurately) that:
\begin{quote}
When the interaction energy in the third dimension vanishes,
Onsager’s exact solution of the 2D Ising model is recovered
immediately.
{\em This guarantees the correctness of the exact solution
of the 3D Ising model} [emphasis added].
\end{quote}
In fact, there is no such guarantee. 
Condition 2 is a necessary but not a sufficient condition for
a solution to be correct.  
For example, the expression (for $v_i$ as defined before)
\bea
\ln[Z]  &=& \ln [2 \cosh[K_1] \cosh[K_2] \cosh[K_3]]
\nonumber \\
& & 
+ \frac{1}{2\,(2\, \pi)^3} \, 
\int_{-\pi}^{\pi} dk_1 \, dk_2 \, dk_3
\nonumber \\
& & 
\times
\ln\bigg[(1+v^2_1) \, (1+v^2_2) \, (1+v_3^2) 
\nonumber \\  
& &
  - 2 \, v_1 \, (1-v_2^2) \, (1-v_3^2) \cos[k_1]
  \nonumber \\
  & & 
  - 2 \, v_2 \, (1-v_1^2) \, (1-v_3^2) \cos[k_2] 
  \nonumber \\
  & & 
  - 2 \, v_3 \, (1-v_2^2) \, (1-v_1^2) \cos[k_3] 
  \bigg],
\eea
correctly reduces to Onsager's solutions of the 2D model
if any one of the three  $K_i$ are made to vanish. 
But this expression clearly is not correct because it 
violates Conditions 1, 3, 4 and 6 above.
See also the famous (but wrong) conjecture of Mark Kac, 
e.g., in Ref.~\cite{feynman-page149}.

\section{Final remarks and conclusion}

The last few decades have witness significant developments
\cite{istrail-2000,o01,3d-kac-ward} aiming to obtain an
exact expression for the 3D Ising model.  
In the absence of strong theoretical results pointing
otherwise, such steady progress should dispel the false 
myth regarding the (non)solvability of the 3D model
(see below).

First, we emphasize that the ferromagnetic 3D Ising model with nearest
neighbor interactions is {\it not} a NP-complete problem.  It is true
that there is a theorem concerning NP-completeness due to Sorin
Istrail~\cite{istrail-2000}.  Nonetheless, it refers to the 3D Ising
spin glass with arbitrary interactions, not to the ferromagnetic
model.  Moreover, the problem addressed in \cite{istrail-2000} relates
to finding the ground state.  For the ferromagnetic case, the ground
state is trivial, viz., with all spins aligned (so doubly
degenerated).

Second, although there is strong numerical evidence of
non-solvability of the 3D Ising model in terms of
$D$-finite functions (see Condition 6), mathematical
proofs for this supposition are still lacking.
But if indeed this would be the case, still an exact analytic
solution based on nonholonomic functions could be possible.
Actually, many experts have been careful to make clear that 
the above mentioned  non-solvability of the 3D Ising model is
conjectural. 
Barry Cipra, writing in Science \cite{cipra-2000}, has stated
that 
\begin{quote}
It might still be possible to find exact answers for some 
special cases of the Ising model, Istrail notes. 
In particular, the ferromagnetic case of the 3D Ising 
model may turn out to be simple enough to solve.
\end{quote}

And third, it is also not true that the progress is too slow or that
the problem is hopelessly too difficult.
Nor is it a waste of time
--- quite the contrary.  In the preface to {\it Polygons, Polyominoes
  and Polycubes}, Anthony J. Guttmann writes~\cite{ppp},
\begin{quote}
    This is indeed a golden age for studying such problems. 
    With powerful computers and new algorithms, unimaginable 
    numerical precision in our estimates of properties of many
    of these models is now possible. 
    On the mathematical side, we are developing tools for 
    solving increasingly complex functional equations, while 
    the theory of conformal invariance, and the developments 
    around stochastic Löwner evolution have given us powerful
    tools to predict, and in some cases to prove, new results.
    The scientific community in this field is divided into 
    those who think we will never solve the problem, of say 
    the perimeter or area generating function of self-avoiding 
    polygons in two dimensions, and those who think that we 
    will. 
    I am firmly in the latter camp$\ldots$
\end{quote}

Finally, we briefly discuss other three dimensional lattice systems.
Condition 1 is general and valid for all lattice systems, so long as
the interactions in the Hamiltonian are nearest-neighbor.
In contrast, conditions 2, 4, and 5 as formulated are specific to the
simple cubic lattice.  However, we can expect that there should be
analogs of these conditions for each lattice system. The same should
be true for condition 3 and the numerical value of $K_c$.
Condition 6, however, is the most difficult to generalize. 
Very little is known about the analog of condition 6 for
other lattice systems.

Summarizing, we have reviewed, systematized and enlarged a set of
necessary conditions characterizing a potentially exact $Z$ for
the 3D Ising model.
We have arranged this set into a single framework.
Obviously, this set does not {\it per se} establish a concrete
protocol that can solve the Ising system.
Nevertheless, if even a
single criterion is violated, one can be 100\% certain that the
methodology followed is fatally flawed.
In this sense, the advance reported here has the potential to guide
the maturing of future attempts to obtain the true $Z$.

We emphasize that the discussion about misguided attempts in the
literature presented here by no means has the intention of criticizing
these authors.
Our purpose is solely to illustrate the subtleties and
intricacies of the problem, which has deceived even some of the most
respectable researchers.
Our discussion thus makes clear the real
necessity of clear-cut tests to check the plausibility of
still to come claims. 

Finally, we mention an eventual (although improbable) curious
consequence of our results.
An exact analytic expression for $Z$ should observe
{\em all} the previously addressed requirements.
However, it could be the case that such a function ---
observing the full set of conditions 1--6 --- cannot exist. 
A proof, of course, would settle negatively the possibility
of an analytical $Z$.
But we conjecture that the six conditions are not mutually
inconsistent.\\

\section*{Acknlowledgements}

We thank CNPq (grants 304532/2019-3, 305062/2017-4 and
302051/2018-0), CAPES, and FACEPE for funding. 
MAGP thanks the Brazilian
National Institute of Science and Technology of Complex Systems
for funding.
GMV thanks V.M.~Kenkre for discussions about boundary conditions
during his stay at the Consortium of the Americas for
Interdisciplinary Science, at the University of New Mexico, during
2008-2009.


\begin{thebibliography}{mt1}

\bibitem{Gershenfeld-1998}
N. Gershenfeld
{\em The Nature of Mathematical Modeling}
(Cambridge University Press, Cambridge, 1998).

\bibitem{borwein-2013}
J. M. Borwein and R. E. Crandall,
Closed forms: what they are and why we care,
Notices Am. Math. Soc. {\bf 60} (1), 50-65 (2013).


\bibitem{ising1924}
E. Ising, Z.  Physik {\bf 31}, 253 (1925).

  
\bibitem{tranquillo}
J. Tranquillo,
{\it An Introduction to Complex Systems: Making Sense of a Changing World}
(Springer, 2019).
  

\bibitem{ons}  
L. Onsager,
{Crystal Statistics. I. A Two-Dimensional 
Model with an Order-Disorder Transition},
{Phys. Rev.} {\bf 65}, 117 (1944).

\bibitem{ruelle-1999}
D. Ruelle,
{\em Statistical Mechanics: Exact Results}
(World Scientific, Singapore, 1999).

\bibitem{thompson-2015}
C. J. Thompson,
{\em Mathematical Statistical Mechanics}
(Princeton University Press, Princeton, 2015).

\bibitem{sutherland-2005}
B. Sutherland,
{\em Beautiful Models: 70 Years of Exactly Solved Quantum
Many-Body Problems}
(World Scientific, Singapore, 2005).


\bibitem{er1} D. C. Mattis, {\it The Many-body Problem: An Encyclopedia of
  Exactly Solved Models in One Dimension} (World Scientific, Singapore,
  1993).

  
\bibitem{er2}
  E. H. Lieb and D. C. Mattis,
  {\it Mathematical Physics in One Dimension} (Academic Press Inc., New York, 1966).


\bibitem{singh-2020}
S. P. Singh,
``The Ising Model: Brief Introduction and Its Application'', 
in S. Sivasankaran, P. K. Nayak, E. Günay (eds.)
{\em Solid State Physics - Metastable, Spintronics 
Materials and Mechanics of Deformable Bodies - Recent Progress}, 
(IntechOpen, London, 2020),
10.5772/intechopen.90875.

\bibitem{maddox2}
J. D. van der Waals,
{\it On the Continuity of the Gaseous and Liquid States},
J. S. Rowlinson (ed.)
(Dover, Mineola, 2004),
%

\bibitem{maddox1}
J. R. Maddox,
talk presented at StatPhys 2 in Paris, 1952.

\bibitem{maddox3}
J. R. Maddox,
{\it Changement de Phases}
Societé de Chimie Physique
(Presses Universitaires de France, Paris, 1952).

\bibitem{perk-comment}
J. H. H. Perk, 
Comment on `Conjectures on exact solution of three-dimensional
(3D) simple orthorhombic Ising lattices',
Philos. Mag. {\bf 89}, 761-764 (2009).

\bibitem{das}  
D. D. Das,
Indian J. Phys. {\bf 44}, 244 (1970).

\bibitem{lou}  
S. L. Lou and S. H. Wu,
Three-dimensional Ising model and transfer matrices,
Chinese J. Phys. {\bf 38}, 841-854 (2000).

\bibitem{zd-zhang}  
Z.-D. Zhang,
Conjectures on the exact solution of three-dimensional
(3D) simple orthorhombic Ising lattices,
Philos. Mag. {\bf 87}, 5309–5419 (2007).
%

\bibitem{degang}
Degang Zhang, 
Exact solution for three-Dimensional Ising model,
Symmetry {\bf 13}, 1837 (2021).
%

%

\bibitem{majewski-2001}
J. Majewski, H. Li, J. Ott,
The Ising model in physics and statistical genetics,  
Am. J. Hum. Gen. {\bf 69}, 853-862 (2001).

\bibitem{binek-2003}
C. Binek,
{\em Ising-type Antiferromagnets}
(Springer, Berlin, 2003).

\bibitem{suzuki-2013}
S. Suzuki, J.-I. Inoue, B. K. Chakrabarti,
{\em Quantum Ising Phases and Transitions in Transverse
Ising Models, 2nd Ed.}
(Springer, Berlin, 2013).

\bibitem{mccoy-2014}
B. M. McCoy and T. T. Wu,
{\em The Two-Dimensional Ising Model, 2nd Ed.}
(Dover, Mineola, 2014).

\bibitem{adler-2016}
M. Adler,
{\em Monte Carlo Simulations of the Ising Model}
(Anchor Academic Publishing, Hamburg, 2016).

\bibitem{fadil-2020}
Z. Fadil,
{\em Semi-infinite Ising model by the Renormalization
Group: Applicable in nanotechnology and spintronics}
(Lap Lambert Academic Publishing, Beau Bassin, 2020).

\bibitem{farah-2020}
A. Farah,
The applications of the Ising model in statistical
thermodynamics and quantum mechanics,
Eur. Acad. Res. {\bf 8}, 2229-2237 (2020).

\bibitem{lipowski-2022}
A. Lipowski (Ed.),  
Special Issue ``Ising Model: Recent Developments and Exotic
Applications'',
Entropy (2022).

\bibitem{cuevas-2016}
G. D. las Cuevas and T. S. Cubitt,  
Simple universal models capture all classical spin physics,
Science {\bf 351}, 1180-1183 (2016).

\bibitem{polyakov-1987}
A. M. Polyakov,
{\em Gauge Fields and Strings}
(Harwood Academic Publishers, London, 1987).

\bibitem{showk-2012}
S. E.-Showk, M. F. Paulos, D. Poland, S. Rychkov, D. S.-Duffin,
A. Vichi,
Solving the 3D Ising model with the conformal bootstrap,
Phys. Rev. D {\bf 86}, 025022 (2012).
  
\bibitem{rychkov-2020}
S. Rychkov,  
3D Ising model: a view from the conformal bootstrap island,
Comptes Rendus. Physique {\bf 21}, 185-198 (2020).

%

\bibitem{b-c-1}
M.-C. Wu and C.-K. Hu,
Exact partition functions of the Ising model on
$M \times N$ planar lattices with periodic–aperiodic
boundary conditions,
J. Phys. A {\bf 35}, 5189-5206 (2002).

\bibitem{b-c-2}
N. Sh. Izmailian and C.-K. Hu,
Finite-size effects for the Ising model on helical tori,
Phys. Rev. E {\bf 76}, 041118 (2007).

\bibitem{cipra-2000}
B. Cipra,
Statistical physicists phase out a dream,
Science {\bf 288}, 1561-1562 (2000).

\bibitem{ruelle-1974}
D. Ruelle,
{\em Statistical Mechanics: Rigorous Results}
(World Scientific, Singapore, 1999).

\bibitem{velenik-2017}
S. Friedli and Y. Velenik,
{\em Statistical Mechanics of Lattice Systems:
  A Concrete Mathematical Introduction} 
(Cambridge University Press, Cambridge, 2017).

\bibitem{feynman-page149}
R. P. Feynman, 
{\em Statistical Mechanics: A Set Of Lectures}
(CRC Press, Boca Raton, 1998).

\bibitem{gattringer}  
C. R. Gattringer, S. Jaimungal, G. W. Semenoff,
Loops, surfaces and Grassmann representation in
two- and three- dimensional Ising models,
Int. J. Mod. Phys. A {\bf 14}, 4549-4574 (1999).
%
       
\bibitem{ott_2020}  
S. Ott,
Weak mixing and analyticity of the pressure in the Ising model,
Commun. Math. Phys. {\bf 377}, 675-696 (2020).

%
%
%
%
%
  
\bibitem{ott_2020_preprint}
S. Ott,
Weak mixing and analyticity in Random Cluster and low temperature
Ising models,
arXiv:2003.05879
 
\bibitem{mardsen-1998}
J. E. Marsden and M. J. Hoffman,
{\em Basic Complex Analysis, 3nd Ed.}
(W. H. Freeman, New York, 1998).

\bibitem{krantz-2002}
S. G. Krantz and H. R. Parks,
{\em A Primer of Real Analytic Functions, 2nd Ed.} 
(Birkh\"auser, Boston, 2002).

\bibitem{beta-c}
A. L. Talapov and H. W. J. Bl\"ote,
The magnetization of the 3D Ising model, 
J. Phys. A {\bf 29}, 5727-5733 (1996).
%

\bibitem{blote-1996}
H. W. J. Bl\"ote, J. R. Heringa, A. Hoogland, E. W. Meyer,
and T. S. Smit,
Monte Carlo renormalization of the 3D  Ising model: 
Analyticity and convergence,
Phys. Rev. Lett. {\bf 76}, 2613-2616 (1996).

\bibitem{landau2018}
A. M. Ferrenberg, J. Xu, D. P. Landau,
Pushing the limits of Monte Carlo simulations for the
three-dimensional Ising model,
Phys. Rev. E {\bf 97}, 043301 (2018).

\bibitem{perk-comment2}
J. H. H. Perk,
Comment on ``Exact Solution for Three-Dimensional Ising
model'' by Degang Zhang,
arXiv:2202.03136.
                
\bibitem{guttmann-enting}
A. J. Guttmann and I. G. Enting,
Series studies of the Potts model. I: The simple cubic Ising
model,
J. Phys. A {\bf 26}, 807-822 (1993).
%

\bibitem  {enting-de-neef}
T. de Neef and I. G. Enting,
Series expansions from the finite lattice method,
J. Phys. A {\bf 10} 801-806 (1977).

\bibitem  {jensen2000}
I. Jensen,
Size and area of square lattice polygons,
J. Phys. A {\bf 33}, 3533-3544 (2000).

\bibitem  {clisby}
N. Clisby,
Enumerative combinatorics of lattice polymers,
%
Not. Am. Math. Soc. {\bf 68} (04), 504-515 (2021).
%

\bibitem{niculescu-2006}
C. P. Niculescu and L.-E. Persson,
{\em Convex Functions and Their Applications, 2nd ed.}
(Springer, Cham, 2018).
        
\bibitem{hansel}
D. Hansel, J. M. Maillard, J. Oitmaa, and M. J. Velgakis,
Analytical properties of the anisotropic cubic Ising model,
J. Stat. Phys. {\bf 48}, 69-80 (1987).

\bibitem{guttmann-prl}
A. J. Guttmann and I. G. Enting,
Solvability of some statistical mechanical systems,
Phys. Rev. Lett. {\bf 76}, 344 (1996).

\bibitem{hoven-2001}
J. Van Der Hoven,
Fast evaluation of holonomic functions near and in
regular singularities,
J. Symb. Comput. {\bf 31}, 717-743 (2001).

\bibitem{fisher-comment}
F. Y. Wu, B. M. McCoy, M. E. Fisher, L. Chayes,
Comment on a recent conjectured solution of the
three-dimensional Ising model,
Phil. Mag. {\bf 88}, 3093-3095 (2008).
	
\bibitem{rosengren}
A. Rosengren,
On the combinatorial solution of the Ising model,  
J. Phys. A. {\bf 19}, 1709-1714 (1986).

\bibitem{istrail-2000}
S. Istrail,
Statistical mechanics, three-dimensionality and
NP-completeness: I. Universality of intractability
for the partition function of the Ising model across
non-planar lattices,
in {\em STOC '00: Proceedings of the Thirty-Second Annual
ACM Symposium on Theory of Computing}
F. Yao, E. Luks (Eds.)
(ACM Press, New York, 2000), pp. 87-96.
%

\bibitem{o01}
T. Regge and R. Zecchina,
Combinatorial and topological approach to the 3D Ising model,
J. Phys. A {\bf 33}, 741-761 (2000).
  
\bibitem{3d-kac-ward}
D. Cimasoni,
A generalized Kac-Ward formula,
J. Stat. Mech. P07023 (2010).
\bibitem{ppp}
A. J. Guttman (Ed.)
{\em Polygons, Polyominoes and Polycubes}
(Springer, Dordrecht, 2009).



\end{thebibliography}
\end{document}